\newcommand{\be}{\begin{equation}}
\newcommand{\ee}{\end{equation}}
\newcommand{\bea}{\begin{eqnarray}}
\newcommand{\eea}{\end{eqnarray}}
\newcommand{\tr}{{\rm tr}\;}
\newcommand{\bi}{\bibitem}
\documentstyle[11pt]{article}
\def\ft#1#2{{\textstyle{{\scriptstyle #1}\over {\scriptstyle #2}}}}

\def\a{\alpha}
\newcommand{\eqn}[1]{eq.\,(\ref{#1})}
\def\eqns#1#2{eqs.\,(\ref{#1}-\ref{#2})}

\renewcommand{\section}[1]{\addtocounter{section}{1}
       \vspace{5mm} \par \noindent
        {\bf \thesection . #1}\setcounter{subsection}{0}
        \par \vspace{2mm} } 
\renewcommand{\subsection}[1]{\addtocounter{subsection}{1}
         \vspace{2.5mm}\par\noindent {\em \thesubsection . #1}\par
          \vspace{0.5mm} }
\renewcommand{\thebibliography}[1]{
          { \vspace{5mm}\par\noindent
          {\bf References}\par\vspace{4mm} } \list{[\arabic{enumi}]}    
          {\settowidth\labelwidth{[#1]}\leftmargin
          \labelwidth \advance\leftmargin\labelsep\addtolength{\topsep}{-4em}
          \usecounter{enumi}}  }

\begin{document}
\pagestyle{empty}
\rightline{UG-2/96}
\rightline{CTP TAMU-8/96}
\rightline{ENSLAPP-A-572/96} 
\rightline{hep-th/9605087}
\vspace{1truecm}
\centerline{\bf COUPLINGS OF SELF-DUAL TENSOR MULTIPLET IN SIX DIMENSIONS}
\vspace{1truecm}
\centerline{\bf E.~Bergshoeff}
\centerline{Institute for Theoretical Physics}
\centerline{Nijenborgh 4, 9747 AG Groningen}
\centerline{The Netherlands}
\vspace{.5truecm}
\centerline{\bf E.~Sezgin\footnote{
Supported in part by the U.S. National Science Foundation, under
grant PHY--9411543.}}
\centerline{Center for Theoretical Physics, Texas A\&M University,}
\centerline{College Station, Texas 77843--4242, U.S.A.}
\vspace{.5truecm}
\centerline{and}
\vspace{.5truecm}
\centerline{\bf E.~Sokatchev}
\centerline{Laboratoire d'Annecy-le-Vieux de Physique des Particules}
\centerline{Bo\^{\i}te postale 110, F-74019 Annecy-le-Vieux Cedex, France}
\vspace{1.3truecm}
\centerline{ABSTRACT}
\vspace{.5truecm}
The $(1,0)$ supersymmetry in six dimensions admits a tensor multiplet which
contains a second-rank antisymmetric tensor field  with a self-dual field
strength and a dilaton. We describe the fully supersymmetric coupling of 
this multiplet to Yang-Mills multiplet, in the absence of supergravity. The
self-duality equation for the tensor field involves a Chern-Simons modified 
field strength, the gauge fermions, and an arbitrary dimensionful parameter. 
\vfill\eject
\pagestyle{plain}
 
\section{\bf Introduction}
\vspace{3mm}

In a spacetime of Lorenzian signature, $p$--forms with self-dual field strengths
can occur in $2\ mod\ 4$ dimensions.  Thus, restricting our
attention to dimensions $D\le 11$, a scalar field in $D=2$, an antisymmetric
tensor in $D=6$ and a four-form potential in $D=10$ can have self-dual field
strengths. Let us refer to these fields as chiral $p$--forms. Chiral scalars
have been extensively studied in the context of world-sheet string actions. The
chiral four-form arises in Type IIB supergravity in $D=10$. The field equations
of this theory have been worked out \cite{IIB}, and are known to be
anomaly-free \cite{AGW}. 

	The remaining supermultiplets which contain chiral $p$--forms exist
in $D=6$. The $(1,0)$ supersymmetry admits the following 
multiplets of this kind: \footnote{For a collection of reprints in which a
large class of
supermultiplets and their couplings are described, see \cite{SS}.} 
\bea
&&{\rm (1,0)\ Supergravity:}\quad\quad\quad   (g_{\mu\nu}, \psi_\mu^i, 
B_{\mu\nu}^- )\ , \nonumber\\
&&{\rm (1,0)\ Matter:}\quad\quad\quad\quad\quad\quad  (B_{\mu\nu}^+, \chi^i,
      \phi)\ , \label{1}
\eea
where $i=1,2$ is an $Sp(1)$ index, and $B_{\mu\nu}^{-}$ and $B_{\mu\nu}^{+}$
are the chiral
two-form potentials with (anti) self-dual field strengths. The $(2,0)$
supersymmetry, on the other hand, admits the following two multiplets with
chiral two-forms:
\bea
&&{\rm (2,0)\ Supergravity:}\quad\quad\quad   (g_{\mu\nu}, \psi_\mu^i, 
B_{\mu\nu}^{ij-} )\ , \nonumber\\
&&{\rm (2,0)\ Matter:}\quad\quad\quad\quad\quad\quad  (B_{\mu\nu}^+, \chi^i,
\phi^{ij})\ , \label{2}
\eea
where $i=1,...,4$ is an $Sp(2)$ index, and $B_{\mu\nu}^{ij-}$, $\phi^{ij}$ are
in the 5-plet representations of $Sp(2)$. 

	There exist also supermultiplets of $(2,1)$, $(3,0)$, $(3,1)$ and
$(4,0)$ supersymmetry in six dimensions that contain chiral two-forms \cite{S},
but these are rather strange multiplets whose field theoretic realizations are
unknown, and we shall not consider them any further in this paper. 

In the case of $(2,0)$ supersymmetry, the equations of motion describing the
coupling of $n$ tensor multiplets to supergravity have been constructed
\cite{Romans1}. The only anomaly-free coupling occurs when $n=21$
\cite{PKT}, in which case the chiral two-forms transform as an $26$-plet of
a global $SO(5,21)$ and the scalar fields parametrize the coset
$SO(5,21)/SO(5)\times SO(21)$.  As was shown in
\cite{PKT}, this model corresponds to Type IIB supergravity compactified on K3.

In the case of $(1,0)$ supersymmetry, one can show that an anomaly-free
coupling of any number of tensor multiplets to supergravity is not
possible. In fact, considering the coupling of supergravity to $n$
tensor multiplets, $V$ vector multiplets and $H$ hypermultiplets, the
necessary but not sufficient condition for anomaly freedom is that
$H-V+29n=273$ \cite{SS}. (This condition must be satisfied to cancel the
${\rm tr} R^4$ terms in the anomaly polynomial). Anomaly-free
combinations of multiplets that arise from certain compactifications of
anomaly-free $N=1, D=10$ supergravity plus Yang-Mills system on $K3$
have been considered in \cite{GSW}. Other anomaly free combinations,
whose $D=10$ origins (if any) are unknown, have been found in
\cite{SS1}\footnote{Witten
\cite{Witten/3} has discovered a new mechanism by which a
nonperturbative symmetry enhancement occurs, and a new class of
anomaly-free models, not realized in perturbative string theory, emerge
in six dimensions. Schwarz \cite{Schwarz1} has
constructed new anomaly-free models in six dimensions, some of which
may potentially arise in a similar nonperturbative scheme.}. 

Rather general couplings of the $(1,0)$ supergravity multiplet to a
single tensor matter multiplet plus an arbitrary number of Yang-Mills
and hypermultiplets have been constructed \cite{ns1}. In this case, the
self-dual and anti-self-dual tensor fields combine to give a single
field strength without any self-duality conditions. In fact, all the
anomaly-free models discussed in \cite{GSW} are of this type. The only
self-dual couplings that are known so far are the following: (i) pure
self-dual supergravity \cite{Pure6D}, (ii) $n$ tensor multiplets ($n>1$)
to supergravity \cite{Romans2}, and (iii) coupling of $n$ tensor
multiplets ($n>1$) {\it and} Yang-Mills multiplets to supergravity
\cite{Sagnotti} \footnote{We are grateful to Edward Witten for bringing
this paper into our attention.}. The $(1,0)$ supergravity by itself is
anomalous, but a systematic analysis of anomalies is required when
tensor and Yang-Mills multiplets are coupled. In particular, a
generalized form of Green-Schwarz anomaly cancellation mechanism in
which a combined action of all the antisymmetric tensor fields has to be
taken into account was shown to apply in this case \cite{Sagnotti}. 

In this paper, we will especially focus on the coupling of self-dual
tensor multiplet to Yang-Mills. One of our motivations for considering
this system is the fact it may play a significant role in the physics of
tensionless strings that have emerged in $M$-theory compactifications to
six dimensions \cite{tensionless}. Moreover, a self-dual string of
the type discussed recently in \cite{Schwarz2} may also exist with
$(1,0)$ supersymmetric anomaly-free coupling to the tensor plus
Yang-Mills system.

Another motivation for considering the self-dual tensor multiplet
couplings in six dimension is that they may play a role in the
description of the dynamics of a class of super $p$--branes. In fact,
the $(2,0)$ tensor multiplet arises as a multiplet of zero-modes
\cite{GT} for the five-brane soliton of \cite{Guven}. As for the $(1,0)$
tensor multiplet, it is natural to look for a super five-brane soliton
in seven dimension, whose translational zero modes would be described by
the dilaton field contained in this multiplet. In fact, a super
five-brane soliton in seven dimensions has been found \cite{lpss1}.
Although the nature of the zero-mode multiplet for this soliton has not
been established, due to a peculiar asymptotic behavior, it seems
plausible that it is actually the self-dual tensor multiplet \cite{lpss1}. 

Matter-modified self-duality equtions in six dimensions may also be
useful in developing a further understanding of the electric-magnetic 
duality symmetry of matter coupled $N=2$ supersymmetric Yang-Mills system, in
a fashion described in \cite{Ve1} for the purely bosonic case.

Finally, matter-modified self-duality equations, known as the ``monopole
equations'' \cite{Wi1}, in the context of a topological Yang-Mills plus
hyper-matter system in $D=4$ \cite{Fre} have also appeared in the
literature. These equations, among other things, have led to important
developments in the study of Donaldson invariants of four-manifolds. One
may ask the question if these equations have a six
dimensional origin as well.

Given the above considerations, we are motivated to consider new types
of interactions of the self-dual tensor multiplet in $D=6$. We have
indeed found that the self-dual tensor multiplet can consistently be
coupled to Yang-Mills multiplet. To our best knowledge, this coupling
has not been noted before in the literature. Of course, the coupling of
self-duality condition-free tensor multiplet to Yang-Mills are known to
occur in supergravity plus Yang-Mills systems in various dimensions,
including $D=6$. However, one can not simply take the flat spacetime
limit to generate the coupling of the tensor field to the Yang-Mills
field, because the latter couples to the former via Chern-Simons form
which is proportional to the gravitational coupling constant. The
novelty of the construction in this paper is the consideration of an
arbitrary dimensionful coupling constant, and the construction of the
interacting self-dual tensor multiplet plus Yang-Mills system directly
by a Noether procedure, without any reference to supergravity. In
Sec. 4 of this paper, we shall comment further on this point and
speculate about a possible mechanisms that might yield an interacting
global limit of the supergravity models constructed in
\cite{ns1,Sagnotti}.

The tensor plus Yang-Mills system considered here exhibits supersymmetry
even when the Yang-Mills system is off-shell, while the tensor multiplet
is on-shell. In trying to put the Yang-Mills sector on-shell, we have
encountered the following surprizing phenomenon: While the tensor field
equations involve the coupling of Yang-Mills system, the latter obey the
free field equations! We explain this phenomenon by writing down an
action for the coupled system in superspace that involves a Lagrange
multiplier superfield that imposes the self-duality condition, but
otherwise decouples from the tensor plus Yang-Mills system. We also show
how this works in component formalism.

In Sec. 2 we will briefly recall the superspace construction of the pure
anti self-dual $(1,0)$ supergravity, and the pure self-dual tensor
multiplet equations. As a side remark, we will show why the coupling of only
Yang-Mills to $(1,0)$ supergravity is impossible. We will then proceed
to a detailed description of the main result of this paper, namely the
coupling of self-dual tensor multiplet to Yang-Mills. Here, we shall
also discuss the phenomenon of free supersymmetric Yang-Mills equations
being consistent with self-dual tensor field equations involving
Yang-Mills supermultiplet. In Sec. 3, we will show how the superspace
constraints of various self-dual systems considered in this paper are
consistent with the $\kappa$-symmetry of the Green-Schwarz superstring
in $D=6$. We summarize our results in Sec. 4, which also contains
further comments on the issue of flat spacetime limit of matter coupled
$(1,0)$ supergravity in $D=6$, and gauge anomalies in the
self-dual tensor plus Yang-Mills system considered in this paper.
 
\section{Self--dual supergravity, tensor multiplet and tensor
multiplet coupled to Yang-Mills} 

We begin by considering an $(1,0)$ superspace in $D=6$ with coordinates
$Z^M=(X^\mu, \theta^{\alpha i})$ where $\theta^{\alpha i}$  are symplectic
Majorana-Weyl
spinors carrying the $Sp(1)$ doublet index $i=1,2$. The basic superfields we
shall consider are the supervielbein $E_M{}^A$, the super two-form $B={1\over
2!} dZ^M\wedge dZ^N\, B_{NM}$ and the Lie algebra valued Yang-Mills super
one-form $A=dZ^M\, A_M$. (Our conventions for super $p$-forms are as in 
\cite{Howe}). Next we define the torsion super two-form $T^A$, the
super three-form $H$ and the Yang-Mills curvature two-form $F$:
\be
T^A=dE^A\ ,\quad\quad H=dB\ ,\quad\quad F=dA+A\wedge A\ ,\label{defs}
\ee
which satisfy the following Bianchi identities
\be
dT^A=E^B\wedge R_B{}^A\ ,\quad\quad dH=0\ , \quad\quad DF=0\ ,\label{Bianchis}
\ee
where $R_B{}^A$ is the Riemann curvature two-form and $D=d+A$. Next, we briefly
review the superspace constraints which describe the on-shell pure
supergravity and pure tensor multiplets.

\subsection{\bf Pure anti-self-dual supergravity}

With the Yang-Mills fields $A$ set to zero, the appropriate torsion and
curvature constraints that describe
the on-shell pure $(1,0)$ supergravity theory in $D=6$ are given by \cite{Be1}
\bea
&& T_{\alpha i,\beta j}{}^a=2\Gamma^a_{\alpha\beta}\epsilon_{ij}\ ,\nonumber\\
&&T_{\alpha i, b}{}^c =0\ ,\quad T_{\alpha i, b}{}^{\gamma k}=0\ , 
\quad T_{\alpha i,\beta j}{}^{\gamma k}=0\ , \nonumber \\
&& H_{a \alpha i,\beta j}= -2 (\Gamma_a)_{\alpha \beta}\epsilon_{ij}\ ,
\label{puresg}\\
&& H_{ab\alpha i}=0\ ,\quad H_{\alpha i,\beta j,\gamma k}=0\ , 
\nonumber \\
&& H_{abc}^{-}=T_{abc}\ ,\quad  H^+_{abc}=0\ ,\nonumber
\eea
where $H_{abc}^{-}$ is anti--self--dual projected and 
$H^+_{abc}$ is  self--dual projected, i.e. $H^\pm_{abc} = 1/2\; (H_{abc} \pm 
\tilde H_{abc}$). For an explicit description of the resulting field equations,
we refer the reader to \cite{Pure6D,Romans2}. 

\subsection{\bf Anti-self-dual supergravity plus Yang-Mills?}

We next consider the
coupling of pure anti--self--dual supergravity to Yang--Mills, and show that an
inconsistency arises. To this end, let us first define a Chern-Simons modified
super three-form ${\cal H}$ as follows \cite{Du2,Ne1,Ka1}:
\be
{\cal H} = \ft12 dZ^M dZ^N dZ^P \left(\partial_P B_{NM}-{\a'\over 2}\, \tr
(A_P F_{NM} 
-{2\over 3} A_P A_N A_M )\right) \ ,
\label{HCS}
\ee
where $\a'$ is an arbitrary dimensionful constant. This three-form satisfies
the Bianchi identity
\be
d{\cal H}={\a'\over 8}\, {\rm tr}\, F\wedge F \ . \label{MBI}
\ee
To couple Yang-Mills to supergravity, we may impose the constraints (\ref
{puresg}), with the replacement $H\rightarrow {\cal H}$ everywhere, and in
addition we impose the {\it off-shell} super-Yang-Mills constraint
\be
F_{\alpha i,\beta j}=0\ .\label{YMC}
\ee
The Bianchi identity $DF=0$ is then  solved, as usual, by setting
\be 
F_{a\alpha i}=-(\Gamma_a)_{\alpha\beta} W^{\beta}_i\ ,\label{YMS}
\ee
where $W_i^\beta$ is a chiral spinor superfield whose leading component is
the gauge
multiplet fermion. Further, the Bianchi identities imply the following
structure of the spinor derivative
\be\label{conseq}
D^i_\alpha W^{\beta j} = \delta^\beta_\alpha Y^{ij} + \epsilon^{ij}
F^\beta_\alpha \; .
\ee
Here $Y^{ij}$ (symmetric in $i,j$) and $F^\beta_\alpha$ (traceless in
$\alpha, \beta$) are superfields whose leading components are the auxiliary
fields and the Yang-Mills field strength, correspondingly. 

To see that the system of constraints described above lead to an
inconsistency, it is sufficient to consider the $(ab,\alpha i,\beta j)$
component of the Bianchi identity (\ref{MBI}):
\bea
&& D_{[a}{\cal H}_{b]\alpha i,\beta j}+D_{(\alpha i} {\cal H}_{\beta j)
ab}+T_{\alpha i,\beta j}{}^C {\cal H}_{C ab} +T_{\alpha i [a}{}^C {\cal
H}_{b]\beta j C}+ T_{ab}{}^C {\cal H}_{C \alpha i,\beta j} \nonumber\\ 
&&\ \ \ \ \ = {3\a'\over 4}\, \tr F_{ab} F_{\alpha i,\beta j} + 
{3\a'\over 4}\, \tr F_{\alpha i[a} F_{b]\beta j} \ . \label{bia}
\eea
We see that as a result of the constraints (\ref{puresg}) and (\ref{YMC}),
the left--hand side vanishes identically, when symmetrized in $i,j$, 
and we are left with the inadmissible equation  $\tr W^\alpha_{(i}
W^\beta_{j)}=0$. 

\subsection{\bf Pure self-dual tensor multiplet}
\vskip 0.3truecm
Again, we begin by setting $A=0$. The pure on-shell $(1,0)$ self-dual tensor
multiplet in $D=6$ is then described by the following superspace constraints
\bea
&& T_{\alpha i,\beta j}{}^a=2\Gamma^a_{\alpha\beta}\epsilon_{ij}\ ,\nonumber \\
&& H_{\alpha i,\beta j,\gamma k}=0\ , \nonumber\\
&& H_{a \alpha i,\beta j}= -2 \phi (\Gamma_a)_{\alpha \beta}\epsilon_{ij}\ ,
\nonumber\\
&& H_{ab\alpha i}=-(\Gamma_{ab})_\alpha{}^\beta\, D_{\beta i}\phi\ ,
\label{puretm}
\eea
with all other components of $T_{AB}{}^C$ vanishing. Here, we have
introduced the dilaton superfield $\phi$. The Bianchi identity
$dH=0$ is now satisfied provided that
\bea
&&H_{abc}^{+}=\Gamma_{abc}^{\alpha\beta} D_{\alpha}^i D_{\beta i}\phi\ , 
\label{c1}\\
&&H_{abc}^{-}=0\ , \label{c2}\\
&& D_\alpha^{(i} D_\beta^{j)} \phi=0\ .\label{c3}  
\eea
In \cite{Ho1}, 
it has been shown that the last constraint describes 
an on--shell self--dual tensor multiplet. 
To see this,
define the physical components of the superfield $\phi$ as follows:
\be
\sigma=\phi|_{\theta=0}\ ,\qquad \chi_{\alpha i}=
D_{\alpha i}\phi|_{\theta=0}\ , \qquad 
H_{abc}^{+}=\Gamma_{abc}^{\alpha\beta} 
D_{\alpha}^{i}D_{\beta i} \phi|_{\theta=0}\ .
\label{onshelltensor}
\ee
Note that the component $H_{abc}^{+}$ in (\ref{onshelltensor}) 
is not, in general, related to the curl of a two--form. 
Then the constraint (\ref{c1}) implies that $H_{abc}^{+}= 
(3\partial_{[a} B_{bc]})^{+}$. 
The constraint (\ref{c2}) is the equation of motion for the 
self--dual tensor field: $(\partial_{[a} B_{bc]})^{-}=0$. In fact, all this 
information, as well as the remaining field equations $\Box \sigma=0$ and 
$\gamma^a \partial_a \chi_{i}=0$, follow from the last constraint (\ref{c3}). 

The quantities appearing in (\ref{onshelltensor}) are field--strengths. 
It is also possible to partially solve  these constraints in terms of gauge 
superfields. To this end we make the following substitution for the 
components of the super--two--form $B$:
\be
B_{\alpha i b} = (\Gamma_b)_{\alpha\beta} V^\beta_i \ , \qquad B_{
\alpha i,\beta j} = 0 \ .\label{choices}
\ee
Inserting this into the constraint \eqns{c1}{c2} we determine the other 
component of $B$,
\be
B_{ab} = (\Gamma_{ab})_\alpha^\beta D_{\beta i} V^{\alpha i} \ ,
\ee
and find an expression for the field--strength $\phi$ in terms of the 
potential $V$:
\be
\phi = D_{\alpha i} V^{\alpha i}\, . 
\label{potential}
\ee
We furthermore derive the constraint
\be
\sigma^{ij}{}^\alpha_\beta \equiv D_{\beta }^{(j} V^{\alpha i)}  - {1\over
4} \delta^\alpha_\beta 
D_{\gamma }^{(j} V^{\gamma i)} = 0\label{constraintpotential}
\ee
on the potential. The latter undergoes gauge transformations which are 
residues of the abelian gauge freedom of the two--form $\delta B = 
d\Lambda$ compatible with the choices (\ref{choices}). 
The constraint (\ref{constraintpotential}) and 
the gauge freedom reduce the content of the superfield $V^{\alpha i}$ to the
potential version of the self--dual tensor multiplet, $\{B_{ab}, \chi^i, 
\sigma\}$, as opposed to the field--strength multiplet $\{H^+_{abc},\chi^i, 
\sigma\}$ described by the superfield $\phi$. It is important to realize
that the left-hand side $\sigma^{ij}{}^\alpha_\beta$ (symmetric in $i,j$
and traceless in $\alpha,\beta$) of eq.
(\ref{constraintpotential}) automatically satisfies the constraint
\be\label{autocon}
D^{(k}_{(\gamma}\sigma^{ij)}{}^\alpha_{\beta)} - {\rm trace} = 0 \; ,
\ee
where $()$ means symmetrization in all the indices involved. This
constraint follows from the spinor derivatives algebra
\be
\{ D^i_\alpha, D^j_\beta\} = 2i \epsilon^{ij}
\partial_{\alpha\beta} \; .  
\ee

\subsection{ \bf Self-dual tensor multiplet coupled to Yang-Mills}

Finally, we consider the
most interesting case of a self--dual tensor  multiplet coupled to Yang-Mills.
Compared to the pure self--dual tensor  multiplet, we need to add the
Yang--Mills field strength $W$. As we already  know, this results in 
Chern-Simons shifts in the three--form $H$. Taking this fact into account, we
propose the following constraints 
\be
{\cal H}_{abc}^{+}=\Gamma_{abc}^{\alpha\beta} 
D_{\alpha}^i D_{\beta i}\phi\ , \label{otm1}
\ee
\be
{\cal H}_{abc}^{-}=\a'(\Gamma_{abc})_{\alpha\beta} \tr W^{\alpha i}
W^\beta_i\ ,
\label{otm2}
\ee
\be
D_\alpha^{(i} D_\beta^{j)} \phi=\a'\epsilon_{\alpha\beta\gamma\delta}
\tr W^{\gamma(i} W^{\delta j)}\ . \label{otm3}
\ee
These constraints are Yang--Mills modified versions of the constraints
(\ref{c1})--(\ref{c3}), and we have shown that they do satisfy the Bianchi
identities (\ref{bia}). We can also use the self--dual tensor multiplet
potential  $V^{\alpha i}$ introduced in (\ref{potential}) to rewrite
eq.~(\ref{otm1})  in the following form (for simplicity we only give the abelian
expression;  the non--abelian generalization is straightforward):
\be\label{constraintt}
D_{\beta }^{(j} V^{\alpha i)}  - {1\over 4} \delta^\alpha_\beta D_\gamma^{(j}
V^{\gamma i)} =  \a' (A^{(j}_\beta W^{\alpha i)} - {1\over 4}
\delta^\alpha_\beta
A^{(j}_\gamma W^{\gamma i)})
{\rm trace}  \ . 
\ee
Clearly, this constraint is a Yang--Mills modified version of the
constraint given in (\ref{constraintpotential}). In it one recognizes
the Chern--Simons type modification due to the Yang--Mills sector. The
reason why such a coupling is consistent can be traced back to the {\it
off-shell} super-Yang-Mills constraint (\ref{YMC}) and its consequence
(\ref{conseq}). Indeed, it is easy to check that the right-hand side of
eq. (\ref{constraintt}) satisfies the same constraint (\ref{autocon}) as
its left-hand side. Note also that the gauge transformation $\delta
A^{j}_\beta = D^{j}_\beta \Lambda$ of the Yang-Mills superfield in
\eqn{constraintt} should be accompanied by the compensating
transformation $\delta V^{\alpha i} = \a' \Lambda W^{\alpha i} $ of the
tensor multiplet potential $V$ (this is typical for Chern--Simons
couplings).

An important point in the above construction is that it requires the
introduction of the dimensionful parameter $\a'$. Although we call it
$\a'$, it is a priori not related to the inverse string tension. It is
natural to expect that this constant gets related to the gravitational
coupling constant or the string tension upon coupling to supergravity.
It is not clear to us, however, how to obtain our results from a
particular flat space limit of the supergravity plus tensor multiplet
plus Yang-Mills system of either \cite{ns1} or \cite{Sagnotti}. We shall
return to this point again in the conclusions section of this paper.

We next show how the above coupling of a tensor multiplet to Yang-Mills
in superspace can be translated to components as well. This can be done
using standard methods. First, the components of the off-shell
Yang--Mills multiplet are contained in the field strength superfield
$W^{\alpha i}$ as follows: 
\bea
\lambda^{\alpha i} = W^{\alpha i}|_{\theta=0}\ , \quad F^{ab} = 
(\Gamma^{ab})^\alpha_\beta D_{\alpha i} W^{\beta i}|_{\theta=0}\  , 
\quad Y^{ij} = D_{\alpha }^{(i} W^{\alpha j)}|_{\theta=0}\  .
\eea
The supersymmetry transformations of these components 
are given by\footnote{
We use the notation and conventions of \cite{Be2}. 
In particular, note that $(A,\lambda, Y_{ij})$ take values in the Lie algebra
of the corresponding gauge group, and that the contraction of $Sp(1)$ indices
in fermionic bilinears is suppressed.}
\begin{eqnarray}
\delta A_a&=& -\bar\epsilon\gamma_a\lambda\, ,\nonumber\\
\delta \lambda^i &=& {1\over 8}\Gamma^{ab} F_{ab} \epsilon^i - {1\over 2}
Y^{ij}\epsilon_j\, ,\\
\delta Y^{ij} &=& - {\bar\epsilon}^{(i}\Gamma^a D_a\lambda^{j)}\, .\nonumber
\end{eqnarray}
The corresponding rules for the on-shell self--dual tensor multiplet
coupled to Yang--Mills are given by

\begin{eqnarray}
\delta \sigma &=& \bar\epsilon\chi\, ,\nonumber\\
\delta\chi^i &=& {1\over 48}\Gamma^{abc} {\cal H}^{+}_{abc}
\epsilon^i +{1\over 4}\Gamma^a\partial_a\sigma\epsilon^i 
- {\a'\over 4}{\rm tr}\,\Gamma^a\lambda^i
\bar\epsilon\Gamma_a\lambda\, ,\label{susy}
\\
\delta B_{ab} &=& -\bar\epsilon\Gamma_{ab}\chi - \a' {\rm tr}\,
A_{[a}\bar\epsilon
\Gamma_{b]}\lambda\, ,\nonumber
\end{eqnarray}
where 

\begin{eqnarray}
{\cal H}_{abc} &=& 3\partial_{[a} B_{bc]} + 3\a'\,{\rm tr}\, 
\left(A_{[a}\partial_b A_{c]}+\ft13 A_a A_b A_c\right) \, ,\nonumber\\
{\cal H}_{abc}^{\pm} &=& {1\over 2}\biggl (
{\cal H}_{abc} \pm \tilde
{\cal H}_{abc}\biggr )\, .
\end{eqnarray}

As in the case of the free self--dual tensor multiplet (\ref{c1})--(\ref{c3}),
it is not hard to see that eqs.~(\ref{otm1})--(\ref{otm3}) 
imply the following 
field equations for the coupled self--dual tensor - Yang--Mills system:

\begin{eqnarray}
{\cal H}_{abc}^{-} &=& -{\a'\over 2}{\rm tr}\,( \bar\lambda
\Gamma_{abc}\lambda)\, , \label{r1}\\
\Gamma^a\partial_a\chi^i &=& \a' {\rm tr}\,\left({1\over 4}\Gamma^{ab} F_{ab}
\lambda^i + Y^{ij}\lambda_j\right)\, ,\label{r2}
\\
\Box \sigma &=& \a' {\rm tr}\,\left(-{1\over 4}F^{ab}F_{ab} - 2\bar\lambda
\Gamma^a
D_a\lambda + Y^{ij}Y_{ij}\right)\, .\label{r3}
\end{eqnarray}
Note that the first constraint leads to the following (dependent)
identity
\be
\partial_{[a}{\cal H}_{bcd]}^+ = \a' {\rm tr}\,\left({3\over 4}
F_{[ab}F_{cd]} -\bar\lambda\Gamma_{[abc}D_{d]}\lambda\right)\, .
\ee

We have verified that the commutator of two supersymmetry
transformations (\ref{susy}) closes on all components of the tensor
multiplet modulo the field equations (\ref{r1})-(\ref{r3}). It is worth
mentioning that \eqn{r1} is already needed for the closure of the
supersymmetry algebra on the tensor field $B$, and \eqn{r2} is needed
for the closure on $\chi$. The last equation can then be derived from
the supersymmetry variation of \eqn{r2}.

The supersymmetry algebra can be expressed as follows:
\be
[\delta (\epsilon_1),\delta (\epsilon_2)] = \delta (\xi^a)+\delta
(\Lambda) + \delta (\Lambda_a)\ , \label{parameters1}
\ee
where the translation parameter $\xi^a$, the tensor gauge transformation 
parameter $\Lambda_a$ and the gauge parameter $\Lambda$ are given by
\be
\xi^a={1\over 2} {\bar\epsilon_2}\Gamma^a\epsilon_1\ ,\qquad
\Lambda_a=\xi^b B_{ba}+\sigma\xi_a\ ,\qquad \Lambda=-\xi^a\Lambda_a\ ,
\label{parameters2}
\ee
and the tensor gauge transformation takes the form
\be
\delta_\Lambda B_{ab}= -{\a'\over 2} \tr \Lambda ( \partial_a A_b -
\partial_b A_a)\ . \label{tgt}
\ee

It should be emphasized that the Yang-Mills system is off-shell, while
the tensor multiplet is on-shell in the coupled system described above.
To put Yang-Mills on-shell, it is natural to impose a condition on the
auxiliary field $Y_{ij}$. Normally, one would set
$Y_{ij}+\chi_{(i}\lambda_{j)}=0$ \cite{Be2}. Here, we encounter a
surprise: The supersymmetric variation of this constraint yields terms
of the type $\a'\epsilon\lambda^3$ that cannot be absorbed into the
field equation of $\lambda$. It turns out that the resolution of this 
problem is to impose the condition
\be
Y^{ij}=0\ . \label{Y}
\ee
This is indeed surprising because it leads to the pure super Yang-Mills
equations
\be
\Gamma^a D_a \lambda=0\ , \quad\quad D_a F^{ab}+
2[{\bar\lambda},\Gamma^b\lambda]=0\ . \label{ymeqs} 
\ee
This peculiarity of the coupling of the on-shell self-dual tensor multiplet
to the off-shell Yang-Mills multiplet is best explained in superspace language.
The {\it on-shell} constraint (\ref{constraintpotential}) of the self-dual
tensor multiplet 
can be obtained from the following action
\be
S=\int d^6x d^8\theta  \ L_{\alpha ij}^\beta[ D_\beta^{(j} V^{\alpha i)} -
{\rm trace}]
 \ , \label{sssaction}
\ee
where $L_{\alpha ij}^\beta$ is a Lagrange multiplier superfield 
symmetric in $ij$ and traceless in $\alpha\beta$. Variation of the action with
respect to this superfield yields the desired constraint
\eqn{constraintpotential}. At the same time, variation of the action with
respect
to $V$ implies
\be
D^j_\beta L_{\alpha ij}^\beta=0\ .\label{leq} 
\ee
This equation propagates the other half (i.e., the anti-self-dual part)
of the tensor multiplet contained in the Lagrange multiplier superfield.
Note that the Lagrange multiplier in the action \eqn{sssaction} has the
gauge invariance
\be\label{gaugeinv}
\delta L_{\alpha ij}^\beta = D^k_\gamma \Lambda_{(ijk)}{}^{(\gamma\beta)}_\alpha
\ee
with parameter $\Lambda$ totally symmetric in $ijk$ and $\gamma\beta$
and traceless in $\alpha\beta$, $\alpha\gamma$. This gauge invariance
corresponds to the 
``conservation law" (\ref{autocon}) of the left-hand side of
\eqn{constraintpotential}. Having written down the free tensor multiplet action
(\ref{sssaction}), we can immediately introduce the
Yang-Mills coupling (\ref{constraintt}) into it:
\be
S=\int d^6x d^8\theta\ L_{\alpha ij}^\beta\left[D_\beta^{(j} V^{\alpha i)}-
\a'A_\beta^{(j} W^{\a i)} - {\rm trace} \right] + \mbox{SYM kin. term}\ ,
\label{ssaction}
\ee
where the last term symbolizes the kinetic term for the super-Yang-Mills
multiplet
\footnote{We shall not need
the explicit form of this kinetic term. Note that in six dimensional
superspace it can
be written in the form of a ``superaction" \cite{Ho1}, $\int d^6x \;
D_{(\alpha i}D_{\beta)}^i
W^{\alpha j} W^{\beta }_j $ which does not involve a Grassmann integral. }.
One needs to make sure that the coupling term is
consistent with the gauge invariance (\ref{gaugeinv}) of the Lagrange
multiplier. This is indeed true, as follows from the argument given after
\eqn{constraintt}. It is very important to realize that this argument only
involves the {\it off-shell} super-Yang-Mills constraint (\ref{YMC}), which
is not modified by the coupling to the tensor multiplet. Clearly, variation
with respect to $L$ of the action (\ref{ssaction}) gives the field
equation of the self-dual tensor multiplet coupled to super-Yang-Mills.
At the same time, variation with respect to $V$ still produces the {\it
free} field equation (\ref{leq}) for the anti-self-dual multiplet.
Finally, the variation of the action with respect to the fields of the
Yang-Mills supermultiplet gives a modification to the free
super-Yang-Mills field equation which is proportional to the Lagrange
multipliers. Now, since $L_{\alpha ij}^\beta$ does not couple to
anything, we can consistently set it equal to zero, once we have derived
all the field equations. This means that on shell the pure
super-Yang-Mills equations are not modified at all.

It is instructive to exhibit the component version of the above
result. First we write the field equations \eqns{r1}{r3} of the
self-dual tensor multiplet in the following form:
\begin{eqnarray}
G^-_{abc} &\equiv & {\cal H}_{abc}^{-} +{\alpha^\prime\over 2}{\rm tr}\,( 
\bar\lambda \Gamma_{abc}\lambda) = 0 \, , \nonumber\\
\Gamma^i &\equiv&
\Gamma^a\partial_a\chi^i -\alpha^\prime {\rm tr}\,\left({1\over 4}\Gamma^{ab} 
F_{ab}
\lambda^i + Y^{ij}\lambda_j\right) =  0\, ,\label{fieldeq}
\\
X &\equiv &\Box \sigma - \alpha^\prime
{\rm tr}\,\left(-{1\over 4}F^{ab}F_{ab} - 2\bar\lambda \Gamma^a
D_a\lambda + Y^{ij}Y_{ij}\right) = 0 \, .
\nonumber
\end{eqnarray}
In showing the supersymmetry of these equations, in effect we have derived the 
following transformation rules:
\begin{eqnarray}
\delta G^-_{abc} &=& - {1\over 2} \bar\epsilon \Gamma_{abc} \Gamma\, 
\nonumber\\
\delta \Gamma^i &=& {1\over 8} \Gamma^{ab} \epsilon^i \partial^c 
G^-_{abc} + {1\over 4} \epsilon^i X\, ,\\
\delta X &=& \bar\epsilon \Gamma^a\partial_a\Gamma\, .\nonumber
\end{eqnarray}
We now introduce a second tensor multiplet with components
$\{\rho,\psi^i,h^+_{abc}\}$ and supersymmetry rules
\begin{eqnarray}
\delta\rho &=& \bar\epsilon\psi\, ,\nonumber\\
\delta \psi^i &=& {1\over 48}\Gamma^{abc}h_{abc}^+\epsilon^i
+ {1\over 4}\Gamma^a\partial_a\rho\epsilon^i\, ,\\
\delta h_{abc}^+ &=& - {1\over 2}\bar\epsilon\Gamma^d\partial_d \Gamma_{abc}
\psi\, .\nonumber
\end{eqnarray}
It is then easy to show that the following Lagrangian is supersymmetric

\begin{equation}
{\cal L}^{(1)} = h^+_{abc} G^-_{abc} + 24 \bar\psi \Gamma
- 6\rho X\, .
\end{equation}
Note that the equation of motion for $B_{ab}$ reads $\partial^a h_{abc}^+$, 
which implies that $h_{abc}$ is a field strength for a potential
$C_{ab}$, namely $h_{abc}=3\partial_{[a} C_{bc]}$.

The SYM Lagrangian, which is separately supersymmetric is given by

\begin{equation}
{\cal L}^{(2)} = \alpha^\prime 
{\rm tr}\,\left(-{1\over 4}F^{ab}F_{ab} - 2\bar\lambda \Gamma^a
D_a\lambda + Y^{ij}Y_{ij}\right)\, .
\end{equation}
The Lagrangian ${\cal L} = {\cal L}^{(1)} + {\cal L}^{(2)}$ describes
the supersymmetric tensor plus Yang-Mills coupled system. Since we have
already shown that the total Lagrangian is supersymmetric, the
supersymmetric Yang-Mills field equations are guaranteed to transform
into each other. These equations are determined by the following
on-shell equation for the auxiliary scalars $Y^{ij}$:
\begin{equation}
\bigl (1+6\rho \bigr ) Y^{ij} = -12 \bar\psi^{(i}\lambda^{j)}\, .
\end{equation}
Strictly speaking we have {\it two} tensor multiplets
coupled to SYM (the Lagrange multipliers are propagating)
\footnote{This is not surprizing, since it is well known that
actions for self-dual fields can only be written with the help of
propagating Lagrange multipliers \cite{MS}. }. 
The second tensor multiplet can be consistently set equal to zero,
however, and that yields the results derived earlier by superspace
methods, namely \eqn{Y}, \eqn{ymeqs} and \eqns{r1}{r3}.

\section{\bf Six-Dimensional Superstring in Self-Dual
                 Backgrounds}

In this section we will show that the $\kappa$ symmetry of the six-dimensional
Green-Schwarz superstring is consistent with the backgrounds described above. 
The action, including the coupling of background non-Abelian Yang-Mills field,
is given by
\bea 
S&=&\int d^2 \xi \big[ -{1\over 2}\phi \sqrt{-g}g^{mn}E_m^a E_n^a
+{1\over 2}\epsilon^{mn}\partial_m Z^M \partial_n Z^N B_{NM} \nonumber\\
&-& {\a'\over 2}(\sqrt{-g} g^{mn}+\rho^{mn})\tr J_m J_n \label{string}
\\
&+&\a'\epsilon^{mn}
(\tr \partial_m y^I L_I\partial_n Z^M A_M +{1\over 2}\partial_m y^I 
\partial_n y^J b_{IJ})\big]\ .\nonumber  
\eea
Here, $\xi^m=(\tau,\sigma)$ are the world-sheet coordinates, $g_{mn}$
is the world-sheet metric and $E_m^a=\partial_m Z^M(\xi) E_M^a(Z)$. 
The field $\rho^{mn}(\xi)$ 
is a Lagrange multiplier whose role is to make the group coordinate
bosons chiral \cite{Si1}. It satisfies the condition
$\rho^{mn} = P^{mp}_+ P^{nq}_+\rho_{pq}$, 
where $P_{+mn}= {1\over 2} (g_{mn} +\sqrt{-g} \epsilon_{mn})$ 
is the projector for self-duality on the world sheet.
The Lie algebra valued one-form
\be
J_m = \partial_m y^I L_I - \partial_m Z^M A_M
\ee
contains the group vielbeins $L_I(y)$. The curl of the two-form $b_{IJ}(y)$
gives the structure constants of the group $G$. 

The $\kappa$--symmetry transformation rules are given by \cite{Ka1}
\begin{eqnarray}
&& \delta Z^M E_M^a = 0\ , 
\nonumber \\
&& \delta Z^M E_M^{\alpha i} =  \Gamma^{\alpha\beta}_a E^a_m \epsilon^{ij}
P^{mn}_+\kappa_{n,\beta j}\ , 
\nonumber \\
&& \delta y^I L_I= \delta Z^M A_M \ , \label{kappa}   
\\
&& \delta\rho^{mn}=-\delta \bigl (\sqrt{-g} g^{mn} \bigr )\ ,  \nonumber \\
&& \delta(\sqrt{-g}g^{mn}) =  2\sqrt{-g} P^{mp}_+P^{nq}_+
\left[-2E_p^{\alpha i} 
+ E_p^a(-u^{\alpha i}_a + \Gamma_a^{\alpha\beta} h^i_\beta)\right.  \nonumber \\
&& \left. \ \ \ -2\a'\phi^{-1}   (2\sqrt{-g} g_{pr} + 
\rho_{pr} )\tr (J^r W^{\alpha i}) \right]\kappa_{q,\alpha i}\ . \nonumber
\end{eqnarray}
Here $\kappa_{m,\alpha i}(\xi)$ is the transformation parameter and 
$u^{\alpha i}_a(Z)$ and $h_{\alpha i}(Z)$ are arbitrary superfields \cite{Se1}.

The invariance of the action (\ref{string}) under the $\kappa$--symmetry 
transformations (\ref{kappa}) imposes the following constraints on the 
background superfields \cite{Se1}
\bea
&&T_{\alpha i\beta j}{}^c =2(\Gamma^c)_{\alpha\beta}\epsilon_{ij}\ ,\quad
T_{\alpha i(bc)} =u^\beta{}_{i(b}\Gamma_{c)\beta\alpha}+
\eta_{bc} (h_{\alpha i} - {1\over 2} \phi^{-1}D_{\alpha i}\phi )
\ ,\nonumber\\
&&{\cal H}_{\alpha i\beta j\gamma k} =0\ ,\quad\quad
{\cal H}_{a\alpha i\beta j}=-2\phi(\Gamma_a)_{\alpha\beta}
\epsilon_{ij}\ ,\nonumber\\ 
&&{\cal H}_{ab\alpha i}=
-2\phi(\Gamma_{ab})_\alpha{}^\beta h_{\beta i}
             +2\phi u^\beta{}_{i [a}\Gamma_{b]\beta\alpha} \ ,
\label{constraints}
\\
&&F_{\alpha i\beta j}=0\ ,\qquad\nonumber
F_{a\alpha i}=-(\Gamma_a)_{\alpha\beta} W^{\beta}_i\ . \nonumber
\end{eqnarray}

We now observe that the constraints (\ref{puresg}), which describe pure
anti-self-dual supergravity, are consistent with the $\kappa$--symmetry
constraints (\ref{constraints}). To see this, we set  $\phi=1$ and $u_a^{\alpha
i}=h_{\alpha i}=W^{\alpha i}=0$ in (\ref{constraints}). 

We also observe that the constraints (\ref{puretm}), which describe pure
self-dual tensor multiplet, are consistent with the $\kappa$--symmetry 
constraints (\ref{constraints}). To see this, we set $u_a^{\alpha i}=W^{\alpha
i}=0$ and $h_{\alpha i}={1\over 2} \phi^{-1} D_{\alpha i} \phi$. 

Finally, to see that the self-dual tensor multiplet coupled to Yang-Mills is
consistent with the $\kappa$--symmetry constraints (\ref{constraints}), we
set  $u_a^{\alpha i}=0$ and $h_{\alpha i}={1\over 2} \phi^{-1} D_{\alpha i}
\phi$ in (\ref{constraints}). 
 
\section{\bf Conclusions}

In this paper we have constructed the coupling of self-dual tensor
multiplet to Yang-Mills in six dimensions. This result is surprising in
the sense that common experience teaches us that Yang--Mills
Chern--Simons terms usually occur only when a supergravity system is
coupled to a matter multiplet. The dimensionful parameter in front of
the Chern--Simons term is then proportional to the gravitational
coupling constant $\kappa$ and, when gravity is turned off, the
Chern--Simons coupling disappears. This phenomenon is somewhat
reminiscent, however, of globally supersymmetric sigma models in four
dimensions which contain the dimensionful scalar self coupling constant
$F_\pi$. At least, in the case of $N=1$ supersymmetric sigma models, it
is known that $F_\pi$ gets quantized in units of the gravitational
coupling constant $\kappa$, upon coupling to supergravity \cite{sigma1}.
Interestingly enough, this relation does not always occur, as was
pointed out by Bagger and Witten \cite{sigma2}, who showed that scalar
self-couplings allowed in global $N=2$ supersymmetry are forbidden in
supergravity, and vice versa. Assuming that the latter case does not
occur in our model, one may expect that the a priori arbitrary
dimensionful coupling constant $\a'$ may indeed get related to $\kappa$
upon coupling to supergravity, or to the inverse string tension $\a'$,
in its dual formulation. Nonetheless, as mentioned earlier, it is not
clear to us at present how to obtain our results from a flat space limit
of any known matter coupled $D=6$ supergravity theory. It is conceivable
that certain stringy constants that arise in the model of
\cite{Sagnotti}, which are essentially undetermined by supersymmetry, may
play a role in defining the global limit seeked. 

It would also be interesting to see whether there is a natural
interpretation of our dimensionful parameter within the context of a
tensionless string in six dimensions \cite{tensionless}, or a super
five-brane theory whose world-volume degrees of freedom would coincide
with those described in our model.

An interesting feature of the tensor--Yang-Mills coupling we constructed
in this letter is that the self-duality condition for the antisymmetric
tensor (see \eqn{r1}) is modified by the Yang--Mills sector. To be
precise it contains the following two contributions from the Yang--Mills
sector: (i) the definition of ${\cal H}$ contains a Yang--Mills
Chern--Simons term and (ii) the right-hand-side of the self-duality
condition contains a bilinear in the Yang--Mills fermions. Such
Yang--Mills modified self-duality conditions are reminiscent of the
monopole equations occurring in \cite{Wi1}. Another potentially
interesting connection is that certain properties of electro--magnetic
duality of Maxwell's theory in four dimensions can be naturally
understood by regarding the theory as a dimensional reduction of a
self-dual tensor in six dimensions \cite{Ve1}. 

In this paper, we have also shown that (a) the coupling of Yang-Mills to
pure anti-self-dual supergravity is not possible, (b) the constraints
describing pure anti-self-dual supergravity or self-dual tensor
multiplet, or coupled self-dual tensor multiplet plus Yang-Mills system
are consistent with the constraints that are imposed by the
$\kappa$--symmetry of the six-dimensional Green-Schwarz superstring
action, and (c) the surprizing phenomenon that while the tensor field
equations involve the coupling of Yang-Mills system, the latter obey the
free field equations. 

We conclude with a remark on anomalies in the self-dual tensor plus
Yang-Mills system considered in this paper. The only possible local
anomaly is the gauge anomaly due to the minimal coupling of the
Yang-Mills field with the chiral gauge fermions. The anomaly polynomial
is thus proportional to $({\rm dim\ G})\,\tr F^4$. The associated gauge
anomaly can be cancelled by Green-Schwarz mechanism provided that the
anomaly polynomial factorizes as $(\tr F^2)^2$. A shown by Okubo
\cite{Okubo}, this factorization is possible only for the gauge groups
$E_8, E_7, E_6, F_4, G_2, SU(3), SU(2), U(1)$, or any of their products
with each other.

\bigskip

\noindent {\bf Acknowledgements}
\vspace{.5truecm}

We thank M.J. Duff, J. Liu, H. Nishino and E. Witten for stimulating
discussions. The work of E.B.~has been made possible by a fellowship of
the Royal Netherlands Academy of Arts and Sciences (KNAW). E. Se. and E.
So. would like to thank the Institute for Theoretical Physics in
Groningen for hospitality. 

\pagebreak

\end{document}